\documentclass[12pt]{article}
\usepackage{amsmath,amssymb,array}
\usepackage{graphicx}

\numberwithin{equation}{section}

\def \p{\partial}
\def \nn{\nonumber}

\begin{document}

\begin{titlepage}

\begin{flushright}
hep-th/0611115
\end{flushright}
\vspace{1.5cm}

\begin{center}

\textbf{\Large{A Note on q-Deformed Two-Dimensional Yang-Mills  \\[0.5cm]
                      and Open Topological Strings}}          \\[1.5cm]

\large{Peng Zhang} \\[1cm]

\emph{Institute of Theoretical Physics, Chinese Academy of Science, \\
      P.O. Box 2735, Beijing 100080, P. R. China} \\[0.2cm]

\textsf{pzhang@itp.ac.cn}

\end{center}
\vspace{1cm}

\centerline{\textbf{Abstract}}\vspace{0.5cm}

In this note we make a test of the open topological string version
of the OSV conjecture, proposed in hep-th/0504054, in the toric
Calabi-Yau manifold $X=\,O(-3)\rightarrow\mathbf{P}^2$ with
background D4-branes wrapped on Lagrangian submanifolds. The D-brane
partition function reduces to an expectation value of some inserted
operators of a q-deformed Yang-Mills theory living on a chain of
$\mathbf{P}^1$'s in the base $\mathbf{P}^2$ of $X$. At large $N$
this partition function can be written as a sum over squares of
chiral blocks, which are related to the open topological string
amplitudes in the local $\mathbf{P}^2$ geometry with branes at both
the outer and inner edges of the toric diagram. This is in agreement
with the conjecture.

\end{titlepage}

\setcounter{footnote}{0}


\section{Introduction}
Microstate counting of four-dimensional BPS black holes, arising
from compactifications of type II superstring theory on Calabi-Yau
3-folds, has received significant progress in recent years.
Ooguri, Strominger and Vafa \cite{OSV} proposed a remarkable
relation between this counting problem with the topological string
theory. They argued that the mixed ensemble partition function
$Z^\mathrm{BH}$ of BPS black holes is related to the topological
string amplitude $\psi^\mathrm{top}$ in the same 3-fold as
\begin{eqnarray}\label{OSV}
Z^\mathrm{BH}=\,|\psi^\mathrm{top}|\,^2  \,\,,
\end{eqnarray}
to all orders of the 't Hooft $1/N$ expansion. Since the left hand
side has definite sense even for finite $N$, the OSV relation (\ref{OSV})
can also be viewed as a non-perturbative completion of the corresponding
topological string theory.

This conjecture has been tested in \cite{V2d, AOSV} by realizing BPS
black holes through D-branes wrapped on cycles in a special class of
Calabi-Yau manifolds which is a vector bundle of rank 2 on a Riemann
surface $\Sigma_g$ with genus $g$. In this case the black hole partition
function can be effectively obtained from a q-deformed two-dimensional
Yang-Mills defined on $\Sigma_g$. Due to the noncompactness of these Calabi-Yau
spaces the original relation (\ref{OSV}) is modified by an additional
summation over an integer, which is interpreted as measuring the
Ramond-Ramond fluxes through $\Sigma_g$, and the appearance of ghost branes,
which encode the extra closed string moduli (of the physical type II theory)
at infinity. In \cite{ANV} the OSV relation is extended
to open topological strings, which capture the information of BPS states
involving some ``background'' D-branes wrapped on certain Lagrangian
submanifolds of the Calabi-Yau 3-folds. In \cite{AJS} the OSV is tested in more
general toric geometries, see also \cite{Pest, Jaff, Sza3, Inst, Sza4}.
Another direction of development is the study of large $N$ phase transition
of the q-deformed 2-d Yang-Mills on the sphere, see e.g.
\cite{Mar1, JM, Sza1, Sza2, Mar2}.
Remarkably, the authors of \cite{Strom1} give a physical argument why
the OSV relation is true from the point of view of M-theory and the
$\mathrm{AdS}_3/\mathrm{CFT}_2$ correspondence. Some related works
can be found in \cite{Dijk1, Strom2, Strom3, Dijk2}.

In this short note we will make a test of the open topological string version
of the OSV conjecture, proposed in \cite{ANV}, in the toric Calabi-Yau manifold
\begin{eqnarray}
X\,=\,\,O(-3)\rightarrow\mathbf{P}^2
\end{eqnarray}
with background D4-branes wrapped on proper Lagrangian submanifolds.
The organization of this note is as follows: In section 2 we review
this open topological string version of OSV. In section 3 we explain
the brane configurations and construct the corresponding partition function.
We wrap three stacks of $N$ D4-branes on three toric divisors respectively,
and also three stacks of $M\,(M<N)$ ``background'' D4-branes on three
Lagrangian submanifolds. Besides degrees of freedom on each branes, there
are additional bifundamental matter fields localized on the intersections
of these branes. To engineering the D0 and D2 charges we need to turn on
some proper Chern-Simons couplings. The partition function of this brane
system can be effectively reduce to a q-deformed two-dimensional Yang-Mills
theory living on a necklace of $\mathbf{P}^1$'s in the base $\mathbf{P}^2$ of $X$.
The matter arising from to the intersections of two divisors reduces to a
point operator, from the two-dimensional viewpoint, inserted at the point
where two $\mathbf{P}^1$'s meet together. While the matter due to the
intersection of the divisors and the Lagragians corresponds to a so-called
holonomy freezing operator on each of three $\mathbf{P}^1$'s. The partition
function we needed can be constructed by gluing some basic building-block
amplitudes properly. In section 4 we study the large $N$ factorization of
the partition function and related it to the open topological string amplitude
with ``background'' D-branes sitting on the inner edges of the toric diagram
and ghost branes on the outer edges. This result is completely in agreement
with the conjecture proposed in \cite{ANV}.


\section{Open topological string version of OSV}
In \cite{ANV} the original OSV relation is extended to open topological strings.
Besides the D4-D2-D0 branes wrapped various cycles to engineer black hole charges,
they also introduce additional ``background'' D4-branes which wrap Lagrangian 3-cycles
and fill a 1+1 dimensional subspace of the Minkowski spacetime. The background branes
induce a new gauge theory in this 1+1 dimensions. Now the brane system contain more
BPS microstates. The electric charges with respect to the 1+1 dimensional gauge field
are open D2-branes ending on the Lagrangian 3-cycles, while the magnetic charges are
domain walls in the 1+1 dimensional theory. Having taken these facts into account
the authors of \cite{ANV} propose that, just as the original OSV, the open topological
string amplitude also captures the chiral sector of the brane partition function at
large $N$.

In their paper this proposal is checked on a particular class of Calabi-Yau 3-folds
\begin{eqnarray}
X\,=\,\,O(-p)\oplus O(p-2)\rightarrow\mathbf{P}^1\,.
\end{eqnarray}
On the divisor $\mathcal{D}=O(-p)\rightarrow\mathbf{P}^1$ there are $N$ D4-branes wrapped.
The D2 and D0 charges are introduced by turning on proper Chern-Simons couplings. The
Lagrangian 3-cycle $\mathcal{L}$, where $M\,(M<N)$ background D4-branes wrapped, is
chosen such that $\mathcal{L}$ meets $\mathcal{D}$ along a circle $\gamma$ . There are
additional bifundamental matter fields localized along $\gamma$. It is argued that,
from the point of view of the D4-branes on $\mathcal{D}$, the only effect of these
matter fields is that the first $M$ eigenvalues of the holonomy along $\gamma$ are
fixed. The theory on $\mathcal{D}$ can be reduced to a q-deformed Yan-Mills
on the base $\mathbf{P}^1$. Suppose $\gamma$ lies entirely on $\mathbf{P}^1$
the effect of background branes is the insertion of the operator
$\delta_M\left(e^{i\oint_\gamma A},\,e^{i\phi}\right)$, where $A$ is the
two-dimensional gauge connection on $\mathbf{P}^1$. However the is an ambiguity
due to the partition of $p$ pinched points on $\mathbf{P}^1$ to two sides of $\gamma$,
i.e. a choice of two integers $p_1$ and $p_2$ such that $p_1+p_2=p$.
They find that this fact corresponds to the framing ambiguity in the open topological
string description. Eventually they gain the large $N$ factorization of the brane
partition function and identify the chiral blocks with the corresponding open
topological string amplitudes as
\begin{eqnarray}
Z_{\mathrm{YM}}(N,g_s,\theta,\phi)&=&\sum_{l\in\mathbb{Z}}\,
   \int\hspace{-0.2cm}\int d_{\mathrm{H}}\phi_1'\,d_{\mathrm{H}}\phi_2'\times  \nn\\
&&\hspace{-0.6cm} \psi_{\mathrm{top}}^{\mathrm{g}}(g_s,t+lpg_s,u+lp_1g_s,u')\,
   \overline{\psi_{\mathrm{top}}^{\mathrm{a}}(g_s,t-lpg_s,u-lp_1g_s,u')}\,.
\end{eqnarray}
The parameters are related as
\begin{eqnarray}
t&=&\frac{1}{2}\,g_sN(p-1)-ip\theta\,,\\
u&=&\frac{1}{2}\,g_sN(p_1-1)-i(p_1\theta-\phi)\,,\\
u_1'&=&\frac{1}{2}\,g_sN+i\phi_1'\,,\\
u_2'&=&\frac{1}{2}\,g_sN+i\phi_2'\,.
\end{eqnarray}
The summation over the integer $l$ is interpreted as the RR fluxes through $\mathbf{P}^1$.
The appearance of ghost and anti-ghost branes is due to the physical \emph{closed} string
moduli at infinity. At the level of topological strings either open or closed string
viewpoint is available. In the physical type II theory, however, only the closed one remains.


\section{Branes in $\,O(-3)\rightarrow\mathbf{P}^2$}

The toric Calabi-Yau 3-fold $X=O(-3)\rightarrow\mathbf{P}^2$ can be described,
in the language of the two dimensional $\mathcal{N}=2$ gauged linear sigma model,
by four chiral fields $X_\mu\,(\mu=0,1,2,3)$, with $U(1)$ charges $(-3,1,1,1)$, as
\begin{eqnarray}\label{P2}
|X_1|^2+|X_2|^2+|X_3|^2-3|X_0|^2=t\,.
\end{eqnarray}
This equation is just the D-flatness condition of the GLSM, which defines a submanifold
$\tilde{X}$ in $\mathbb{C}^4$. The 3-fold $X$ is defined by the quotient of $\tilde{X}$
mod by the action of the gauge group $U(1)$. It is a holomorphic line bundle of degree $-3$
over the base $\mathbf{P}^2$ which is represented by $X_0=0$. It is a toric manifold,
and its toric diagram is
\begin{center}
\includegraphics{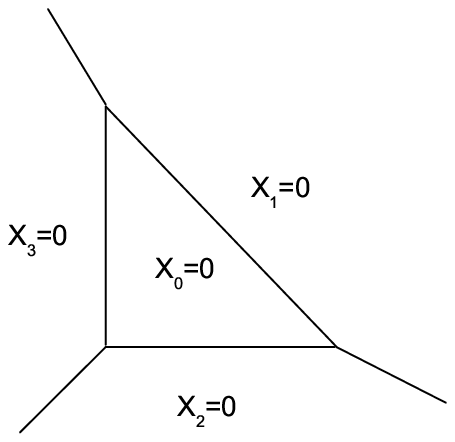}
\end{center}

We wrap $N$ D4-branes respectively on the divisors $\mathcal{D}_i$ which is defined, in
addition to to (\ref{P2}), by the equations
\begin{eqnarray}\label{Di}
\mathcal{D}_i: \quad X_i=0,\quad i=1,2,3.
\end{eqnarray}
These submanifolds are just the total space of a line bundle of degree $-3$ over
$\mathbf{P}^1$, i.e.
\begin{eqnarray}
\mathcal{D}_i\,\sim\,\,O(-3)\rightarrow\mathbf{P}^1.
\end{eqnarray}
Any two divisors $\mathcal{D}_i$ and $\mathcal{D}_j$ intersect along a complex plane
specified by $X_i=X_j=0$. There are additional bifundamental matters
localized at these intersections. The base of $\bigcup\mathcal{D}_i$ is a chain of
three $\mathbf{P}^1$'s which meet pairwise at a single point $p_i$.

Since we want to test the relation \cite{ANV} between the \emph{open} topological string
amplitude and the two dimensional Yang-Mills thoery, we also need to add ``background''
D4-branes wrapped on proper Lagrangian submanifolds. The Lagrangian submanifolds of some
simple toric varieties have been studied in e.g. \cite{AV, AKlV}. We choose following three
Lagrangian submanifolds $\mathcal{L}_i$ as
\begin{eqnarray}\label{La}
\mathcal{L}_1:&& |X_1|^2-|X_0|^2=0,\quad |X_2|^2-|X_0|^2=c_1\,. \nn \\
\mathcal{L}_2:&& |X_2|^2-|X_0|^2=0,\quad |X_3|^2-|X_0|^2=c_2\,.     \\
\mathcal{L}_3:&& |X_3|^2-|X_0|^2=0,\quad |X_1|^2-|X_0|^2=c_3\,. \nn
\end{eqnarray}
Here the constants $c_i$ satisfy $0<c_i<t$. Now we wrap $M$ D4-branes on each of these three
Lagrangians. Combine (\ref{Di}) and (\ref{La}) it can be seen that $\mathcal{D}_i$ intersects
with $\mathcal{L}_i$ along a circle $\gamma_i$ which is entirely in three base $\mathbf{P}^1$'s
respectively. As before there are also bifundamental matters along $\gamma_i$.

\subsection{Dynamics on branes}

The low energy dynamics on the D4-branes wrapped on $\mathcal{D}_i=O(-3)\rightarrow\mathbf{P}^1$,
as argued in \cite{V2d, AOSV}, is just the topological twist of the $\mathcal{N}=4$ gauge theory,
studied in \cite{VW}. To turn on the D0 and the D2 charges we need to add the following terms
\begin{eqnarray}
\frac{1}{2g_s}\int_{\mathcal{D}_i}\mathrm{Tr}\,\mathcal{F}\wedge\mathcal{F} +\,
\frac{\theta}{g_s}\int_{\mathcal{D}_i}\mathrm{Tr}\,\mathcal{F}\wedge\mu
\end{eqnarray}
to the original $\mathcal{N}=4$ theory. Here $\mu$ is the area form on $\mathbf{P}^1$.
As discussed in \cite{V2d, AOSV} the theory localized to modes which are invariant under the
$U(1)$ rotation acting on the fiber $\mathbb{C}^1$. This is a crucial observation which effectively
reduces the theory to a two-dimensional gauge theory on $\mathbf{P}^1$.
To see this introduce the variable $\Phi(z)$ by
\begin{eqnarray}
\Phi(z)=\int_{\mathbb{C}^1_z}\mathcal{F}=\oint_{S^1_{z,\infty}}\mathcal{A}\,.
\end{eqnarray}
Now what is needed is the partition function of the theory on $\mathbf{P}^1$ defined by the action
\begin{eqnarray}
S=\frac{1}{g_s}\int_{\mathbf{P}^1}\mathrm{Tr}\,\Phi\,F +\,
  \frac{\theta}{g_s}\int_{\mathbf{P}^1}\mu\,\mathrm{Tr}\,\Phi -\,
  \frac{1}{2g_s}\int_{\mathbf{P}^1}\mu\,\mathrm{Tr}\,\Phi^2\,.
\end{eqnarray}
Integrating out the field $\Phi$ it becomes the standard Yang-Mills action with a $\theta$-term.
However there is a subtlety noticed in \cite{AOSV}: Although the action is just standard, the
integration measure of $\Phi$ is not as usual. It is proper to consider the variable $e^{i\Phi}$,
rather than $\Phi$, as the fundamental field. Taking account of this fact the resulting quantum
theory is the so-called q-deformed two-dimensional Yang-Mils theory.

Because of the intersections of the D4-branes wrapped on $\mathcal{D}_i$ with each other and with
the background D4-branes wrapped on $\mathcal{L}_i$, there should be some additional bifundamental
matter fields localized at these intersections. The effect of integrating out these matter fields,
from the viewpoint of the two-dimensional theory, is the insertion of some operators on the base
sphere $\mathbf{P}^1$. As argued in \cite{AJS}, the matter fields localized on $\mathcal{D}_i\bigcap\mathcal{D}_j$
correspond to  inserting the operator
\begin{eqnarray}\label{V}
\mathcal{V}=\sum_R\, \mathrm{Tr}_R V^{-1}_{(i)}\, \mathrm{Tr}_R V_{(i+1)}\,,
\end{eqnarray}
at the point $p_i$ where two base spheres meet, with
\begin{eqnarray}
V_{(i)}=e^{i\Phi_{(i)}-i\oint A_{(i)}}\,,\quad
V_{(i+1)}=e^{i\Phi_{(i+1)}}\,.
\end{eqnarray}
The integral contour is just a small loop around the meet points.

The divisor $\mathcal{D}_i$ intersects with the Lagrangian submanifold $\mathcal{L}_i$ along a curve $\gamma_i$.
There are also bifundamental matters localized on $\gamma_i$. As discussed in \cite{ANV}, the effect of these
matters, from the point of view of the gauge theory on $\mathcal{D}_i$, is just the insertion of the holonomy
freezing operator $\delta_M\left(e^{i\oint\mathcal{A}},\,e^{i\phi}\right)$, which can be written
in a Weyl invariant way as follows
\begin{eqnarray}
\delta_M\left(e^{i\oint\mathcal{A}},e^{i\phi}\right)=D(\oint\mathcal{A})^{-1}
       \sum_{\sigma\in S_N}(-1)^\sigma\prod_{\alpha=1}^M\,\delta\left((\,e^{i\oint\mathcal{A}})
       _{\sigma(\scriptscriptstyle{\alpha})},\,e^{i\phi_\alpha}\right)\,.
\end{eqnarray}
The insertion of this delta-function type operator fixes the first $M$ eigenvalues of the holonomy
$e^{i\oint\mathcal{A}}$ along $\gamma_i$, which is a $N\times N$ matrix, to be $M$ specified
numbers $e^{i\phi_\alpha}$ with $\alpha=1,\cdots,M$.
Since intersection curve $\gamma_i$ is totally contained in the base $\mathbf{P}^1$, the holonomy
freezing operator $\delta_M\left(e^{i\oint\mathcal{A}},\,e^{i\phi}\right)$ just simply reduce to
\begin{eqnarray}
\delta_M\left(e^{i\oint A},\,e^{i\phi}\right), \label{delta}
\end{eqnarray}
in the two-dimensional theory, where $A$ is the gauge connection on $\mathbf{P}^1$.
As noted in \cite{ANV}, there is an ambiguity which corresponds to the number of the
pinched points on each side of $\gamma_i$, and parametrized by a choice of $a_i$ and $a_i'$
such that $a_i+a_i'=3$, since the degree of the divisor $\mathcal{D}_i$ is $-3$.

Therefore the brane partition function has been reduced to the partition function
of the q-deformed Yang-Mills defined on a chain of three $\mathbf{P}^1$'s, with the observable
(\ref{V}) inserted at each point $p_i$ where two $\mathbf{P}^1$'s meet together,
and the holonomy freezing operator (\ref{delta}) along the curve $\gamma_i$ on each $\mathbf{P}^1$.
\begin{center}
\includegraphics{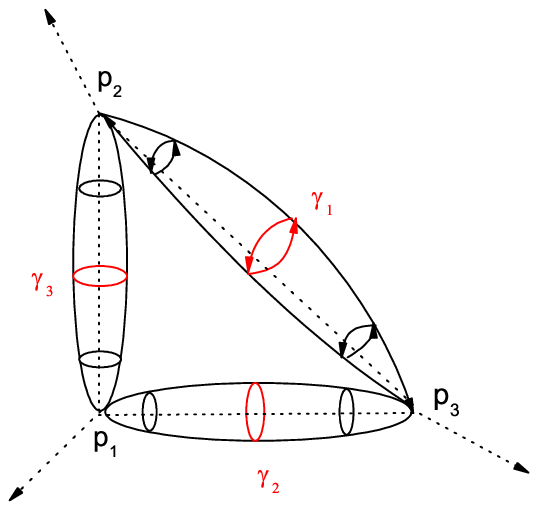}
\end{center}

\subsection{Partition function of 2d qYM}

The q-deformed two-dimensional Yang-Mills theory can be solved exactly by an operatorial approach
\cite{AOSV}. What we need is some basic amplitudes as building blocks. The partition function with
various insertions can be gained by the gluing properly these simple amplitudes. The Hilbert space $\mathcal{H}$
of the q-deformed two-dimensional Yang-Mills can be viewed as the space of class functions on the
gauge group. This is because the path integral over the fields of the theory on a Riemann surface
$\Sigma$ with boundary $\p\Sigma$ gives a function $\Psi(U)$. The group element $U$ is just the holonomy
along the boundary. A convenient basis of $\mathcal{H}$ is the characters $\mathrm{Tr}_R\,U$, which are
also the energy eigenfunctions of the 2-d qYM. These functions satisfy the orthonomality condition
\begin{eqnarray}
\langle\,R\,|\,Q\,\rangle=\delta_{R\,Q}\,.
\end{eqnarray}

To construct the brane partition function we can cut the chain of $\mathbf{P}^1$'s
into several parts: three ``vertexes'' and three annulus with a holonomy freezing operator
inserted. In followings we will review each of these amplitudes.

The partition function on an annulus with area $a$ is \cite{AOSV}
\begin{eqnarray}
\langle{R}|A^{(a)}|{Q}\rangle\,=\,\delta_{RQ}\,e^{-a\left(\frac{1}{2}g_sC_2(R)-i\theta C_1(R)\right)}\,\equiv\,
\delta_{RQ}\,A^{(a)}_R\,,
\end{eqnarray}
where $C_1(R)$ and $C_2(R)$ are respectively the first and  the second Casimir of the representation $R$.
It can be viewed as an ``evolution'' amplitude from state $|Q\rangle$ to state $|R\rangle$
with the Euclidean time parameter $a$. It is zero unless the initial and final states are the same one.

The ``vertex'' amplitude, i.e. the expectation value of the observable (\ref{V}),
has been worked out in \cite{AJS}. It takes the form as
\begin{eqnarray}
\mathcal{V}_{RQ}\,\equiv\,\langle{R}|\mathcal{V}|{Q}\rangle\,=\,
\Theta^N(q)\,S_{\bar{R}Q}(q)\,\,q^{-\frac{1}{2}C_2(R)-\frac{1}{2}C_2(Q)}\,,\quad q=e^{-g_s}\,.
\end{eqnarray}
Here $\Theta^N(q)=\sum_{m\in\mathbb{Z}}q^{m^2/2}$, and
the function $S_{\bar{R}Q}(q)$ is related to the S-matrix of the WZW model through
\begin{eqnarray}
S_{\bar{R}Q}(q)=S_{RQ}(q^{-1})=q^{-\rho^2-\frac{N}{24}}\sum_{\sigma\in S_N}
    (-1)^\sigma q^{\sigma(R+\rho)\cdot(Q+\rho)}\,,
\end{eqnarray}
where $\rho=\frac{1}{2}(N-1,N-3,\cdots,3-N,1-N)$ is the half of the sum of all positive
roots of $U(N)$. The dot in the exponential denotes the standard inner product of $\mathbb{R}^N$,
and $R$ here denotes $(R_1,\cdots,R_N)$ --- number of boxes in each row of the Young diagram.

The expectation value of the holonomy freezing operator is
\begin{eqnarray}\label{holo}
\Delta_{RQ}\,\equiv\,\langle{R}|\,\delta_M\left(e^{i\oint_\gamma A},\,e^{i\phi}\right)|{Q}\rangle\,=\,
\sum_A\,\mathrm{Tr}_{R/A}(e^{-i\phi})\,\mathrm{Tr}_{Q/A}(e^{i\phi})\,.
\end{eqnarray}
Here $\mathrm{Tr}_{R/A}$ is the skew trace, defined by
$\mathrm{Tr}_{R/A}(U)=\sum_Q B^R_{AQ}\,\mathrm{Tr}(U)$
with $B^R_{AQ}$ being the branching coefficients.
More details can be found in \cite{ANV}.

The total partition function can be constructed by the gluing above basic amplitudes as followings
\begin{eqnarray}\label{Z}
Z^{\mathrm{qYM}}(g_s,\theta,\phi_i)&=&Z_0\sum_{R_i,R_i'}
   \mathcal{V}_{R_1'R_2}A_{R_2}^{(a_2)}\Delta_{R_2R_2'}A_{R_2'}^{(a_2')}
   \mathcal{V}_{R_2'R_3}A_{R_3}^{(a_3)}\Delta_{R_3R_3'}A_{R_3'}^{(a_3')}
   \mathcal{V}_{R_3'R_1}A_{R_1}^{(a_1)}\Delta_{R_1R_1'}A_{R_1'}^{(a_1')} \nn\\[0.2cm]
&=&Z_0\,\Theta^{3N}(q)\sum_{R_i,R_i'}S_{\bar{R}_1'R_2}S_{\bar{R}_2'R_3}S_{\bar{R}_3'R_1}
   \Delta_{R_1R_1'}\Delta_{R_2R_2'}\Delta_{R_3R_3'}  \nn\\
&&\hspace{1cm}\times\prod_{i=1}^{3}\,q^{\frac{a_i-1}{2}C_2(R_i)+\frac{a_i'-1}{2}C_2(R_i)}
   \prod_{i=1}^{3}\,e^{i\theta_i\left(a_iC_1(R_i)+a_i'C_1(R_i')\right)}\,.
\end{eqnarray}
There is an ambiguity due to a choice of $a_i$ and $a_i'$ such that $a_i+a_i'=3$.
The normalization factor $Z_0$ is determined by the requirement of the
Large $N$ factorization in the next section.
The first line of (\ref{Z}) can be read off form the following graph.
\begin{center}
\includegraphics{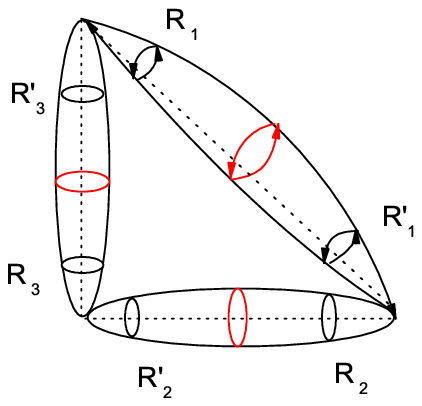}
\end{center}


\section{Large $N$ factorization and open topological strings}
In this section we will study the Large $N$ factorization the the q-deformed 2-d YM into chiral and
anti-chiral sectors, and relate them with the open topological string amplitudes.
To do this an essential technical tool introduced in \cite{GT} is the notion of a
\emph{composite representation}, whose contribution survives in the large $N$ limit.
A composite representation $R$ of $U(N)$ can be labeled by two Young diagrams $R_\pm$ and
a integer $l_R$ as
\begin{eqnarray}
R=R_+\overline{R_-}[\,l_R]\,.
\end{eqnarray}
The meaning of the integer $l_R$ is just the powers of the determinant representation of $U(N)$.
The Casimir operators for a composite representation $R=R_+\overline{R_-}[\,l_R]$ is
\begin{eqnarray}
C_1(R)&=&Nl_R+|R_+|+|R_-|\,,\\
C_2(R)&=&\kappa_{R_+}+\kappa_{R_-}+N(|R_+|+|R_-|)+Nl_R^2+2l_R(|R_+|+|R_-|)\,,\label{C2}
\end{eqnarray}
where $|R|$ denotes the total number of the Young diagram corresponding to the representation $R$, and
\begin{eqnarray}
\kappa_{R}=\sum_{i}R_i(R_i-2i+1)\,,
\end{eqnarray}
which is independent of $N$. The factorization of the $S_{\bar{R}Q}$ has been gained in
\cite{ANV, AJS}
\begin{eqnarray}\label{S fac}
S_{\bar{R}Q}(q)&=&M(q^{-1})\,\eta^N(q^{-1})\,(-q^{\frac{N}{2}})^{|R_+|+|R_-|+|Q_+|+|Q_-|}
   \,q^{l_R(|Q_+|-|Q_-|)+l_Q(|R_+|-|R_-|)}  \nn\\
&&\hspace{0.3cm}q^{Nl_Rl_Q}\,q^{\frac{1}{2}\kappa_{R_+}+\frac{1}{2}\kappa_{R_-}}
   \sum_P(-1)^{|P|} q^{-N|P|} C_{Q_+^tR_+P}(q^{-1})\,C_{Q_-^tR_-P^t}(q^{-1})\,.
\end{eqnarray}
Here $M(q)$ and $\eta(q)$ is the McMahon and Dedekind eta functions,
and $C_{Q_+^tR_+P}(q^{-1})$ is the topological vertex amplitude of \cite{Vertex},
$Q^t$ denotes the transpose representation.
This result can be argued by the geometric transition of the deformed conifold
$\mathrm{T}^*\mathbf{S}^3$ to the resolved conifold $O(-1)\oplus O(-1)\rightarrow\mathbf{P}^1$,
both with four stacks of non-compact D-branes.
The holonomy freezing operator amplitude $\Delta_{RQ}$ in (\ref{holo}) with respect to
two composite representations is \cite{ANV}
\begin{eqnarray}\label{delta2}
\Delta_{RQ}=\delta_{\,l_Rl_Q}\left(\sum_{A_+}s_{R_+/A_+}(e^{-i\phi})\,s_{Q_+/A_+}(e^{i\phi})\right)
    \left(\sum_{A_-}s_{R_-/A_-}(e^{i\phi})\,s_{Q_-/A_-}(e^{-i\phi})\right)\,.
\end{eqnarray}
Here $s_{R/A}(x)$ is the skew Schur function, which is related to the Schur function through
\begin{eqnarray}
s_{R/A}(x)=\sum_Q N^R_{AQ}\,s_Q(x)\,,\label{Schur}
\end{eqnarray}
where the quantity $N^R_{AQ}$ is the Littlewood-Richardson number.
The formula (\ref{delta2}) follows from the relation
between the skew trace and the skew Schur function \cite{ANV}
\begin{eqnarray}
\mathrm{Tr}_{R/A}(e^{i\phi})=\,s_{R_+/A_+}(e^{i\phi})\,s_{R_-/A_-}(e^{-i\phi})\,\,\mathrm{det}e^{i\,l_R\phi}\,.
\end{eqnarray}

Put every things needed into (\ref{Z}) and introduce the parameter $l=l_1+l_2+l_3$.
Up to $O(e^{-N})$ terms, the partition function takes the following factorization form \\
\begin{eqnarray}
Z^{\mathrm{qYM}}(g_s,\theta,\phi_i)=\sum_{l_i\in\mathbb{Z}}\sum_{P_i}(-1)^{|P_1|+|P_2|+|P_3|}\,
   \psi^{\mathrm{top}}_{P_1 P_2 P_3}(g_s,t+lg_s,u_i)\,
   \psi^{\mathrm{top}}_{P_1^t P_2^t P_3^t}(g_s,\bar{t}-lg_s,u_i)\,.\nn\\
\end{eqnarray}
Note that the integers $l_i'$ disappear due to $\delta_{l_il_i'}$ in $\Delta_{R_iR_i'}$.
Here $\psi^{\mathrm{top}}_{P_1 P_2 P_3}(g_s,t,u_i)$ is the open topological string amplitude
with three stacks of ghost branes at each outer edge of the toric diagram, and three
stacks of background branes at each of the inner edges.
The summations over representations $R_{i+}$ and $R_{i+}'$ have been absolved in
$\psi^{\mathrm{top}}_{P_1 P_2 P_3}$, while $R_{i-}$ and $R_{i-}'$ in $\psi^{\mathrm{top}}_{P_1^t P_2^t P_3^t}$.
This result is in complete agreement with the conjecture proposed in \cite{ANV} by
Aganagic, Neitzke and Vafa. In the followings we will give some details.

The parameters of the topological string theory is
\begin{eqnarray}
t&=&\frac{3}{2}\,g_sN-3i\theta\,, \\
u_1&=&\phi_1-a_1\theta-\frac{i}{2}\,a_1\,g_sN-ig_s\left((a_1-1)l_1+l_3\right)\,, \\
u_2&=&\phi_2-a_2\theta-\frac{i}{2}\,a_2\,g_sN-ig_s\left((a_2-1)l_2+l_1\right)\,, \\
u_3&=&\phi_3-a_3\theta-\frac{i}{2}\,a_3\,g_sN-ig_s\left((a_3-1)l_3+l_2\right)\,.
\end{eqnarray}
As discussed in \cite{ANV}, $2\,\mathrm{Re}\,\pi u_i/g_s$ is the chemical potential
corresponding to the additional electric charges due to the Lagrangian branes,
while $\mathrm{Im}\,u_i/\pi$ is the magnetic charges due to the domain walls in 1+1 subspace.
The above relation between parameters can be determined as follows. In the second line
of (\ref{Z}), we choose
\begin{itemize}
\item\quad $q^{\frac{N}{2}(|R'_{1+}|+|R_{2+}|)}$ from $S_{\bar{R}_1'R_2}$,
           and similar terms from $S_{\bar{R}_2'R_3}$ and $S_{\bar{R}_3'R_1}$;
\item\quad $s_{R_{i+}/A_{i+}}(e^{-i\phi_i})\,s_{R'_{i+}/A_{i+}}(e^{i\phi_i})$ from $\Delta_{R_iR_i'}$;
\item\quad $N|R_{i+}|+2l_i|R_{i+}|$ from $C_2(R_i)$, and similar terms in $C_2(R_i')$;
\item\quad $|R_{i+}|$ and $|R'_{i+}|$ from $C_1(R_i)$ and $C_1(R_i')$, respectively.
\end{itemize}
By noting that the skew Schur function $s_{R/A}$ is a homogeneous function with degree $|R|-|A|$,
together with the fact $a_i+a_i'=3$, we can simplify the product of these factors as follows
\begin{eqnarray}
\prod_{i=1}^3 \left(s_{R_{i+}/A_{i+}}(e^{-iu_i})\,s_{R'_{i+}/A_{i+}}(e^{iu_i})\,\,
    e^{-(\frac{3}{2}g_sN-3i\theta_i+lg_s)|R'_{i+}|}\,\right)\,. \label{pro+}
\end{eqnarray}
Here $u_i$'s are defined as above. If requiring three K\"{a}hler moduli
$\frac{3}{2}g_sN-3i\theta_i$ are same, we must have $\theta_1=\theta_2=\theta_3$ denoted by $\theta$,
i.e. $t=\frac{3}{2}\,g_sN-3i\theta$ as above.
If we choose the anti-chiral sectors from above four kind of terms, their product is exactly
\begin{eqnarray}
\prod_{i=1}^3 \left(s_{R_{i-}/A_{i-}}(e^{iu_i})\,s_{R'_{i-}/A_{i-}}(e^{-iu_i})\,\,
    e^{-(\bar{t}-lg_s)|R'_{i-}|}\,\right)\,.\label{pro-}
\end{eqnarray}

In the paper \cite{Vertex} the gluing rule for the A-model topological string amplitude, with brane
sitting on inner edges of the toric diagram, is
\begin{eqnarray}\label{glu}
\sum_{ABB'}(-1)^{s}q^{f}e^{-t|A|}\,C_{PQA\otimes B}\,C_{P'Q'A\otimes B'}\,s_B(e^{-iu})\,s_{B'}(e^{-t+iu})\,.
\end{eqnarray}
Since the topological vertex make sense only for irreducible representations, so
$C_{PQA\otimes B}$ is just the abbreviation for $\sum_A N^R_{AB}\,C_{PQR}$, and $C_{P'Q'A\otimes B'}$ for
$\sum_A N^{R'}_{AB'}\,C_{P'Q'R'}$\footnote{I appreciate M. Marino and A. Neitzke
for the explanation of this point.}. The sign factor $s$ and the frame factor
$f$ only depend on the irreducible representation $R$ and  $R'$. Do the summation over $B$ and $B'$
and note the relation (\ref{Schur}) between the Schur and skew Schur functions, then
(\ref{glu}) becomes
\begin{eqnarray}\label{glu2}
\sum_{ARR'}(-1)^{s(R,R')}q^{f(R,R')}e^{-t|R'|}\,C_{PQR}\,C_{P'Q'R'}\,s_{R/A}(e^{-iu})s_{R'/A}(e^{iu})\,,
\end{eqnarray}
where we have used the homogeneity of the skew Schur functions to extract $e^{-t}$ which cancels
$e^{-t|A|}$ and gives the factor $e^{-t|R'|}$. The structure in (\ref{glu2}) is exactly the same as
what has appeared in the 2d qYM amplitude with holonomy freezing operator $\Delta_{RR'}$ inserted,
\emph{c.f.} (\ref{pro+}) and (\ref{pro-}).

At last we determine the normalization factor $Z_0$. Note that there are quadratic terms of $l_i$ in
the expressions of $C_2(R)$, see (\ref{C2}), and $S_{\bar{R}Q}$, see (\ref{S fac}). The product of
these factors just gives $q^{\frac{1}{2}Nl^2}$ with $l=l_1+l_2+l_3$. This structure is the same as
that of \cite{AJS} although now there are honolomy freezing operators, since we have $a_i+a_i'=3$.
Therefore the normalization factor $Z_0$ is just that of \cite{AJS}, i.e.
\begin{eqnarray}
Z_0=e^{-\frac{3}{8}g_sN^3+\,\frac{\theta^2}{2g_s}N}
\end{eqnarray}
This normalization factor is crucial for the large $N$ factorization and the relation between the
topological string amplitude.

\section*{Acknowledgements}
I would like to appreciate D. Jafferis, M. Marino, A. Neitzke, R. J. Szabo and C. Vafa for
valuable correspondence which give me, as a beginner of this field, much help during the
period of this little work. I also appreciate prof. Miao Li for reading the draft. At last
I would thank Interdisciplinary Center for Theoretical Studies of USTC where this note is completed.

\end{document}